\documentclass[11pt,twoside]{article}
\usepackage{atmp}
\copyrightnotice{2003}{7}{331}{367}   
\newcommand\mathC{\mkern1mu\raise2.2pt\hbox{$\scriptscriptstyle|$}
        {\mkern-7mu\rm C}}          
\newcommand{\mathR}{{\rm I\! R}}         
\newcommand\Q{{\cal Q}}

\newcommand\K{{\cal K}}
\newcommand\id{{\rm id}}
\newcommand\bra[1]{\langle#1|\,}
\newcommand\braket[2]{\langle#1|#2\rangle}
\newcommand\ket[1]{\,|#1\rangle}
\newcommand\Ob[1]{{\rm Ob}(#1)}
\newcommand\Hom[1]{{\rm Hom}(#1)}
\newcommand\Sem[1]{{\rm Sem}(#1)}
\newcommand\Ran[1]{{\rm Ran}\;#1}
\newcommand\Dom[1]{{\rm Dom}\;#1}
\newcommand\AF[1]{{\rm AF}(#1)}
\newcommand\eq[1]{Eq.\ (\ref{#1})}
\newcommand\eqs[2]{Eqs.\ (\ref{#1}--\ref{#2})}

\newcommand\mapright[1]{\smash{
        \mathop{\mbox{\large{$\;\longrightarrow\;$}}}\limits^{#1}}}
\newcommand\mapleft[1]{\smash{
        \mathop{\mbox{\large{$\;\longleftarrow\;$}}}\limits^{#1}}}

\begin{document}
\title{A New Approach to Quantising Space-Time:
        I.\ Quantising on a General Category}
\url{gr-qc/0303060}     
\author{C.J.~Isham}      
\address{The Blackett Laboratory\\
            Imperial College of Science, Technology \& Medicine\\
            South Kensington\\
            London SW7 2BZ}
\addressemail{c.isham@imperial.ac.uk}
\markboth{\it QUANTISING ON A GENERAL CATEGORY\ldots}{\it
C.J.~ISHAM}

\begin{abstract}

A new approach is suggested to the problem of quantising causal
sets, or topologies, or other such models for space-time (or
space). The starting point is the observation that entities of
this type can be regarded as objects in a category whose arrows
are structure-preserving maps. This motivates investigating the
general problem of quantising a system whose `configuration space'
(or history-theory analogue) can be regarded as the set of objects
$\Ob\Q$ in a category $\Q$. In this first of a series of papers,
we study this question in general and develop a scheme based on
constructing an analogue of the group that is used in the
canonical quantisation of a system whose configuration space is a
manifold $Q\simeq G/H$ where $G$ and $H$ are Lie groups. In
particular,  we choose as the analogue of $G$ the monoid of `arrow
fields' on $\Q$. Physically, this means that an arrow between two
objects in the category is viewed as some sort of analogue of
momentum. After finding the `category quantisation monoid', we
show how suitable representations can be constructed using a
bundle of Hilbert spaces over $\Ob\Q$.

\end{abstract}

\cutpage

\section{Introduction}\label{Sec:Introduction}
One of the enduringly, and endearingly, fascinating
challenges in quantum gravity is to give meaning to the
idea of quantising space, or space-time, at a level that
is more fundamental than that of quantising a metric
tensor on a background manifold.\footnote{This is not
meant to imply that it is obvious how to quantise a
metric field on a background space or space-time: it is
not!} For example, one comes across phrases in the
literature such as `quantising causal sets\footnote{A
`causal set' is a partially-ordered set $P$ whose
elements represent points in a discrete model for
space-time. If $p,q\in P$ are such that $p\leq q$ then
the physical interpretation is that an event at $p$ can
causally influence events at $q$.}' \cite{Sor87}
\cite{Sor91} \cite{Sor00} \cite{Rap00}, or `quantising
topology', and the goal of the present paper is to invest
these concepts with a new, and precise, meaning.

In the present paper an operator-based approach to quantising
space, or space-time, structure is described. The ensuing theory
is applicable to two types of physical situation. The first is
`canonical quantisation', where the states in the Hilbert space
refer to the situation at a fixed time. For example, in a theory
of the quantisation of the topology $\tau$ of physical space, the
states might be functions $\tau\mapsto\psi(\tau)$, or an extension
thereof. Such a theory would then need to be augmented with a
`Hamiltonian' operator to specify how these states evolve in
(possibly, a discrete) time.

On the other hand, it is not appropriate to talk about the {\em
canonical\/} quantisation of causal sets since each causal set $c$
is a complete space-time in itself, and hence a state function
$c\mapsto\psi(c)$ has no physical meaning within the
interpretative framework of standard quantum theory. However,
states of this type {\em are\/} meaningful in a consistent-history
approach to quantum theory. More precisely, in the `HPO' (history
projection operator) method, propositions about complete histories
of a system are represented by projection operators on a `history
Hilbert space' \cite{Isham94}. In the case of causal sets, the
propositions would include statements about the causal-set
structure of space-time. In a theory of this type, the analogue of
`dynamics' is coded into the decoherence functional that is to be
constructed from the basic quantum operators in the history
Hilbert space.

It is important to keep in mind these two different ways of using
operators and Hilbert spaces: (i) a canonical quantum theory, and
(ii) the HPO approach to a consistent history theory. The general
mathematical framework is the same in both cases, but the physical
interpretation is quite different.

Let us now consider in more detail the construction a quantum
history theory of causal sets. A first guess might be that the
history state vectors are functions $c\mapsto\psi(c)$, or some
generalisation thereof. This sounds plausible, but how is it to be
justified; and what is the appropriate `generalisation' of a
function $\psi$ of causal sets?

In the canonical quantisation of, say, a particle moving in one
dimension, the rationale for identifying states with wave
functions in the Hilbert space $L^2(\mathR)$ lies in the existence
of operators $\hat x$ and $\hat p$ that are assumed to satisfy the
canonical commutation relations $[\,\hat x,\hat p\,]=i\hbar$. It
then follows from the famous theorem of Stone and Von Neumann that
any irreducible representation of this canonical
algebra\footnote{More precisely, the theorem refers to weakly
continuous representations of the exponentiated form of the
canonical commutation relations.} is unitarily equivalent to the
familiar one on $L^2(\mathR)$ in which $(\hat x
\psi)(x):=x\psi(x)$ and $(\hat p\psi)(x):=-i\hbar d\psi/dx$.

More generally, consider a system whose configuration space is a
finite-dimensional differentiable manifold $Q$ such that $Q\simeq
G/H$, where $G$ and $H$ are Lie groups. The analogue of the
canonical commutation relations includes a representation of the
Lie algebra of $G$, and the elements of this algebra are the
momentum variables in the theory. The question arises, therefore,
of whether there is analogue of momentum for causal sets. When
$Q\simeq G/H$, the Lie group $G$ generates transformations from
one point in $Q$ to another, which leads us to consider how one
causal set can be `transformed' into another\footnote{This
question has also been of particular interest to Rafael Sorkin and
Ioannis Raptis in the context of their own work on causal sets;
private communication.}. Similarly, in quantum topology, we would
seek a natural way of `transforming' from one topological space to
another.

In the case of causal sets, one can imagine trying to remove, or
add, points and links, but it is not easy to describe a general
scheme for doing this. For example, removing a point (and the
associated links) might result in a disconnected causal set. But
suppose we do not wish to admit space-times of this type: what
then? Similarly, adjoining a point to a causal set is not trivial
since enough links must be also added to ensure that the resulting
structure is a partially-ordered set, and this can be done in
different ways.

The key idea of the present paper is that, in the example
of causal sets, what `connects' one causal set to another
is the collection of all order-preserving maps between
them. This suggests that to each such map
$f:c_1\rightarrow c_2$ there is to be associated an
operator $\hat d(f)$. Moreover, if we have three causal
sets $c_1,c_2,c_3$, and order-preserving maps
$f:c_1\rightarrow c_2$, and $g:c_2\rightarrow c_3$, then
the composition $g\circ f:c_1\rightarrow c_3$ is also
order-preserving. It is natural to postulate that the
operators $\hat d(f)$ reflect this structure by
satisfying the relations
\begin{equation}
    \hat d(g\circ f)=\hat d(g)\hat d(f), \label{Intro:main}
\end{equation}
or, perhaps,
\begin{equation}
    \hat d(g\circ f)=\hat d(f)\hat d(g), \label{Intro:mainanti}
\end{equation}
since, at this stage, there is no {\em prima facie\/} reason for
preferring any particular ordering of the operators.

    Note that in a scheme of this type it is
easy to use just connected causal sets, if so desired. Or we can
require the causal sets to be finite; or restrict our attention to
order-preserving maps that are one-to-one; or to any of a number
of variants of the basic idea if there is some good physical
reason for doing so. It is clear that a similar idea could be
applied to topological spaces, with the analogue of
order-preserving maps being continuous functions.

These preliminary ideas could be developed into genuine
quantisation schemes for causal sets and topological spaces.
However, these particular examples admit a natural generalisation
that applies to many different physical situations, and it is this
generalisation that is described in the present paper. The
application of this scheme to causal sets is discussed in a
companion paper \cite{IshQCT2_03}.

To motivate this generalisation, consider the example of, say,
finite causal sets. The key remark is that these can be viewed as
the {\em objects\/} of a category, whose arrows/morphisms are
order-preserving maps. Similarly, one can imagine forming a
category whose objects are topological spaces of some specific
physical interest, and whose arrows are continuous maps.

Thus we are led to the following general problem. Namely,
to construct the quantum theory of a system whose
`configuration space' (or history analogue) is the set of
objects in some category $\Q$, and in which the role of
momentum transformations is played by the arrows in $\Q$.
More precisely, if $f:A\rightarrow B$ is an arrow ({\em
i.e.}, the objects $A$ and $B$ are the domain and range
of $f$ respectively), then we think of $f$ as providing a
(partial) description of how to `transform' from $A$ to
$B$. In general, there will be many arrows from $A$ to
$B$, and we shall regard the set of all of them, denoted
$\Hom{A,B}$, as affording the complete description of how
to transform from $A$ to $B$.

Note that for this idea to be mathematically meaningful,
the category $\Q$ must be `small' in the sense that the
collection of all objects, $\Ob\Q$, in $\Q$,  and the
collection of all arrows, $\Hom\Q$, in $\Q$ must be
genuine sets, not classes. For example, the category of
{\em all\/} sets is certainly not of this type.

To construct a quantum theory on a general small category $\Q$ we
generalise what was said above for causal sets. Thus we expect
each arrow $f\in\Hom\Q$ to be associated with an operator $\hat
d(f)$ in such a way as to represent the law of arrow composition
 by the operator relations \eq{Intro:main} (or
\eq{Intro:mainanti}). If the objects in $\Ob\Q$ have a
physically important internal structure---as manifested
mathematically by the sets $\Hom{A,A}$, $A\in\Ob\Q$,
being non-trivial---this should be reflected in the
quantum theory. In particular, we anticipate that the
(history) state vectors are vector-space valued functions
on $\Ob\Q$, in analogy to what happens for a manifold
$Q\simeq G/H$ where, generically, the states are
cross-sections of a vector bundle over $Q$ whose fibre
carries a representation of $H$. In fact, as we shall
see, the construction of a quantum theory on $\Q$
involves a generalisation of the idea of vector-valued
functions.

A variety of physically interesting situations are
special cases of this categorical scheme. For example:
\begin{enumerate}
\item $\Q$ is a category of finite (perhaps connected)
causal sets interpreted as a history theory. Another
possibility is to use causal sets that are locally
finite. Or one could choose some `master' causal
set\footnote{Fay Dowker: private discussion.} $U$ and let
$\Q$ be the category of all causal subsets of $U$, with
the arrows being the order-preserving embeddings of one
causal set in another.

\item $\Q$ is a small category of
partially ordered sets interpreted canonically as the
structure of physical space at a given `time'.

\item $\Q$ is a small category of topological spaces. This
gives a new approach to `quantum topology':  to be interpreted as
a history theory if the objects represent space-time, and as a
canonical theory if the objects represent space.\footnote{More
generally, $\Q$ could be a small category of locales (in the
context of topology without points) whose arrows are localic
maps.}

\item $\Q$ is a small category of differentiable manifolds, with
the arrows being differentiable maps between manifolds,
regarded as models of either space-time or space.

\item A more bizarre example is to take $\Q$ to be
a small category of groups, with the arrows being group
homomorphisms. For example, perhaps the symmetry group of
a unified field theory undergoes `quantum fluctuations'
near the big bang singularity? This certainly gives a
novel interpretation to the idea of ``quantum group
theory'':-)

\item In all the examples above,
the category $\Q$ is a category of sets with structure, and the
arrows are maps that preserve this structure. Thus it is useful to
look first at the case where $\Q$ is a small category of sets, and
the arrows between two sets $A$ and $B$ are {\em any\/} functions
from $A$ to $B$. This is studied in detail in \cite{IshQCT2_03}.

\item An example of a category whose objects are {\em not\/}
structured sets is a partially-ordered set (poset) $P$. The
objects of this category are the points in $P$, and if $p,q\in P$,
an arrow is defined to exist from $p$ to $q$ if and only if $p\leq
q$ (hence there can be at most one arrow between any two
points/objects). In this case, the objects have no internal
structure, and so the quantum theory should be relatively simple.

This example is useful for providing a mathematically simple
illustration of the general scheme, and we shall discuss it in
\cite{IshQCT2_03}. However, it cannot be interpreted as a theory
of quantum space-time (or space) of the type in which we are
interested since each possible space-time (or space) is
represented by structureless point\footnote{The canonical theory
could be interpreted as that of a particle whose position is
confined to one of the points in the poset $P$, but there is no
obvious reason why such a system should be of any physical
interest.}.

\end{enumerate}

At this point, however, it is important to observe that there is
an obvious problem with imposing \eq{Intro:main} (or
\eq{Intro:mainanti}) as it stands. Namely, the composition $g\circ
f$ is only defined if the range, $\Ran f$, of $f$ is equal to the
domain, $\Dom g$, of $g$. Thus, if $f:A\rightarrow B$, and
$g:C\rightarrow D$, the composition $g\circ f$ is only defined if
$B=C$. On the other hand, the operator product $\hat d(g)\hat
d(f)$ on the right hand side of \eq{Intro:main} (or
\eq{Intro:mainanti}) is always defined\footnote{If the operators
$\hat d(f)$, $f\in\Hom\Q$, are unbounded then these products may
not exist. But that is not the point at issue here which applies
even when the operators concerned are all bounded.}. The
resolution of this issue is one of the key steps in the quantum
scheme. Two approaches are suggested: the first involves a
semigroup $\Sem\Q$ that is constructed from the arrows in $\Q$;
the second, and preferred, method involves the idea of an `arrow
field'.

The plan of the paper is as follows. An initial approach
to developing a quantum theory is discussed in Section
\ref{Sec:QuantisingWithSemQ}. The focus is placed on
equipping the set of arrows $\Hom\Q$ of $\Q$ with a
semigroup structure.  We show how this semigroup,
$\Sem\Q$, generates transformations of the set of objects
$\Ob\Q$: as such, it constitutes our first attempt at
finding an analogue of the group $G$ used in the
quantisation of a system whose configuration space is a
manifold $Q\simeq G/H$.

However, the construction of $\Sem\Q$ is rather coarse,
involving as it does the ad hoc introduction of an
element `$\star$' whose sole role is to serve as the
value of $g\circ f$ when the range of $f$ does {\em
not\/} equal the domain of $g$. This is remedied in
Section \ref{Sec:SemigroupArrowFields} by the
introduction of the idea of an `arrow field', defined to
be an assignment to each object $A\in\Ob\Q$ of an arrow
whose domain is $A$. The crucial property of the set
$\AF\Q$ of all arrow fields is that it has a natural
semigroup structure without the need for additional,
spurious elements.

The action of the semigroup $\AF\Q$ on $\Ob\Q$ is used in Section
\ref{Sec:AFQT} to provide the foundation of the quantum scheme. We
start in Section \ref{SubSec:BasicAlgebraQT} with a simple
approach in which the quantum states are complex-valued functions
on $\Ob\Q$. This suffices to construct the basic `category
quantisation monoid', each of whose faithful, irreducible
representations is deemed to constitute a proper quantisation on
$\Q$.

However, as it stands, this simple scheme is inadequate since the
quantum operators do not distinguish arrows with the same domain
and range. We solve this problem in Section
\ref{SubSec:IntroducingMultiplier} by  generalising the state
vectors to become vector-valued functions on the set of objects
$\Ob\Q$. It transpires that the vector space in which a function
takes its values must vary from object to object, and each such
`multiplier' representation is associated with a presheaf of
Hilbert spaces over $\Ob\Q$. (However, no detailed language of
presheaf theory is used in the present paper.)

The collection of basic quantum operators is completed in Section
\ref{SubSec:Adjointa(X)} with a computation of the adjoints of the
operators $\hat a(X)$ that represent arrow fields $X\in\AF\Q$. In
Section \ref{SubSec:ProdaXadagX}, we compute the products $\hat
a(X)^\dagger\hat a(X)$ and $\hat a(X)\hat a(X)^\dagger$ that might
be expected to play an important role in specific applications of
the quantum scheme. In Section \ref{SubSec:QuestionIrreducibility}
there are a few preliminary remarks about the irreducibility of
the representations we have constructed. The paper concludes with
Section \ref{Sec:Conclusions}, which is mainly a list of problems
for further research.

The present paper, the first in a series, introduces the general
theory of quantising on a category. In the second paper
\cite{IshQCT2_03}, the general theory is developed for the
physically important case where $\Q$ is a category of sets. In
\cite{IshQCT3_03} we return to the general theory and present an
alternative approach in which state vectors are complex-valued
functions on arrows, rather than, as in the present paper, on
objects. Later papers in the series will describe further
developments of some of the main ideas, and more concrete
examples.

\section{Quantising Using the Semigroup ${\rm Sem}({\cal
Q})$}\label{Sec:QuantisingWithSemQ}
\subsection{The Semigroup ${\rm Sem}({\cal
Q})$} The key problem identified above in the context of
\eq{Intro:main} is that the composition $g\circ f$ of arrows is
only defined if $\Ran f =\Dom g$, whereas the operator product
$\hat d(g)\hat d(f)$ (or $\hat d(f)\hat d(g)$) is always defined.
In other words, the set $\Hom\Q$ of arrows is only a {\em
partial\/} semigroup under the law of composing arrows, whereas
the (bounded) operators on a Hilbert space are a full semigroup
under operator multiplication.

In this context, recall that a (full) {\em semigroup\/} is a
non-empty set $S$ with a law of combination that is associative.
Thus
\begin{equation}
       a(bc)=(ab)c          \label{AssocLaw}
\end{equation}
for all $a,b,c\in S$. A semigroup with a unit element is called a
{\em monoid\/}\footnote{This distinction is not very important
since a unit element can always be appended if one is not
present.}. In a monoid (and unlike in a group) elements may not
have inverses.

A {\em partial semigroup\/} $S$ is a more general
structure in which not all pairs of elements $a,b\in S$
can be combined; if a pair $a,b$ {\em can\/} be combined,
they are said to be {\em compatible}. The associativity
law \eq{AssocLaw} is now imposed only when it makes
sense: {\em i.e.}, when the different elements in
\eq{AssocLaw} are compatible in the appropriate way.

For our purposes, a key observation is that the equation
\eq{Intro:main},  $\hat d(g\circ f)=\hat d(g)\hat d(f)$,
{\em would\/} be well defined if the elements $g$ and $f$
belonged to a full semigroup rather than only to a partial
one. Thus, we start by considering if it is possible in
general to convert a partial semigroup into a full
semigroup.

One simple approach is to append an extra element $\star$ to a
partial semigroup $S$, and then try to define a new combination
law `$\&$' by
\begin{equation}
    a\& b:=\left\{ \begin{array}{ll}
     ab &\mbox{if $a$ and $b$ are compatible;} \label{Def:Amp}\\
                 {\star} & \mbox{otherwise,}
                 \end{array}
        \right.
\end{equation}
and
\begin{eqnarray}
        \star\& a &:=& \star                \label{Def:Amp2}\\
        a\& \star &:=& \star                \label{Def:Amp3}\\
        \star\&\star &:=&\star              \label{Def:Amp4}
\end{eqnarray}
for all\footnote{Thus $\star$ is an absorptive element.}
$a\in S$. Note that it follows from these definitions
that if $a$ and $b$ are compatible elements of $S$, then
$a\&b\neq \star$.

The combination law `$\&$' defined by
\eqs{Def:Amp}{Def:Amp4} on the extended set
$S_+:=S\cup\{\star\}$ has the serious failing that, in
general, it is not associative. For example, consider
$a,b,c\in S$ such that $a$ and $b$ are not compatible.
Then, for all $c\in S$, we have
$(a\&b)\&c=\star\&c=\star$. On the other hand, it could
be that $b$ and $c$ {\em are\/} compatible, in which case
$b\&c\neq \star$. Then, if $a$ is compatible with $b\&c$,
which is possible, we have $a\&(b\&c)\neq \star$, and
hence a failure of associativity.

However, this objection does not apply when the partial
semigroup is the set of arrows in a category $\Q$,  with
$a\&b:=a\circ b$ for $a,b\in\Hom\Q$. For $a$ and $b$ are
not compatible if and only if $\Ran b\neq\Dom a$. But if
$b$ and $c$ are compatible, then $b\&c=b\circ c$, and
since $\Ran b\circ c=\Ran b$ it follows that $a$ and $b\&
c$ are not compatible.

In conclusion, the partial semigroup $\Hom\Q$ of arrows in a small
category $\Q$ can be given the structure of a full semigroup by
augmenting the set $\Hom\Q$ with an additional element $\star$
(which is not given a domain or range), and then defining, for all
$f,g\in\Hom\Q$,
\begin{equation}
    g\& f:=\left\{ \begin{array}{ll}
    g\circ f &\mbox{if $\Ran f=\Dom g$;} \label{DefAmp2}\\
                 {\star} & \mbox{otherwise,}
                 \end{array}
        \right.
\end{equation}
and
\begin{eqnarray}
        \star\& f &:=& \star\\
        f\& \star &:=& \star\\
        \star\&\star &:=&\star
\end{eqnarray}
for all $f\in\Hom\Q$. This semigroup will be denoted $\Sem\Q$. It
does not have a unit element.\footnote{We could also construct a
free $\Sem\Q$-algebra over $\mathC$, denoted $\mathC[\Sem\Q]$,
whose elements are defined to be complex-valued functions on
$\Sem\Q$ that vanish for all but a finite number of elements of
$\Sem\Q$ \cite{Lang65}. If $u,v\in\mathC[\Sem\Q]$, their product
$uv$ is defined as $uv(h):=\sum_{f\& g=h}u(f)v(g)$. In the special
case where the category $\Q$ is a partially-ordered set, this
reproduces the incidence algebra used by Raptis and Zapatrin in
their work on discretising space-time topology \cite{Rap00}
\cite{RapZap00} \cite{RapZap01}.}

\subsection{An Embryo Quantum Theory on $\Q$}
If a Lie group $G$ acts on the left on a manifold $Q$ then an
elementary (anti)-representation of $G$ is given on the vector
space of complex-valued functions on $Q$ by $(\hat
U(g)\psi)(q):=\psi(gq)$, and this can be used as a starting point
for discussing the quantum theory of a system whose configuation
space is $Q$ (for more details see Section
\ref{SubSec:QuantManQ}).

As anticipated in Section \ref{Sec:Introduction}, in the category
case, a crucial idea is that the arrows in a category can be
thought of as `transforming' one object into another, which
suggests that perhaps $\Hom\Q$ can play the role of the group $G$
above. Thus, if $f$ is an arrow such that $f:A\rightarrow B$, we
define $\tau_f(A):=B$. However, this leaves open the question of
how to define $\tau_f(A)$ when $A\neq\Dom f$. The simplest way
(although not the one we shall ultimately adopt) is to augment the
set $\Ob\Q$ with an additional element, denoted $\#$, and then to
define the action of an arrow $f$ on the augmented set $\Ob\Q_+$
by
\begin{equation}
\tau_f(A):=\left\{ \begin{array}{ll}
                 \Ran f &\mbox{if $\Dom f=A$;}
                                    \label{Def:taufA}\\
                 {\#} & \mbox{otherwise,}
                 \end{array}
        \right.
\end{equation}
and $\tau_f(\#):= \#$. This can be extended to an action
of the semigroup $\Sem\Q$ by defining $\tau_\star(A):=\#$
for all $A\in\Ob\Q_+$. It is  easy to check that, for all
$f,g\in\Sem\Q$,
\begin{equation}
            \tau_f\circ\tau_g=\tau_{f\&g}.
\end{equation}
Thus, by these means, we have defined a left action of
the semigroup $\Sem\Q$ on the extended set of objects
$\Ob\Q_+$.

This action can be used to give a first attempt at a
quantum theory on $\Q$. The simplest scheme is to choose
state vectors to be complex-valued functions on $\Ob\Q_+$
and then to define operators $\hat d(f)$ by
\begin{equation}
    (\hat d(f)\psi)(A):=\psi(\tau_f(A))\label{Def:df}
\end{equation}
for all $f\in\Sem\Q$ and $A\in\Ob\Q_+$. Then, if $f,g\in\Sem\Q$,
\begin{eqnarray}
    (\hat d(g)\hat d(f)\psi)(A)&=&(\hat
    d(f)\psi)(\tau_g(A))=\psi(\tau_f(\tau_g(A)))\nonumber\\
            &=&\psi(\tau_{f\&g}(A)) =(\hat d(f\&g)\psi)(A)
\end{eqnarray}
and so
\begin{equation}
        \hat d(g)\hat d(f)=\hat d(f\&g)
\end{equation}
which is an (anti)-representation of the semigroup $\Sem\Q$.

One implication of \eq{Def:df} is that if $f\in\Hom\Q$
but $\Dom f\neq A$, then $(\hat d(f)\psi)(A)=\psi(\#)$,
and also $(\hat d(\star)\psi)(A)=\psi(\#)$ for all
$A\in\Ob\Q_+$. In fact, nothing of significance is lost
if we forget the extra element $\#$ in $\Ob\Q_+$ in the
sense that we define the state vectors to be functions on
$\Ob\Q$ only (which is equivalent to setting
$\psi(\#)=0$): {\em i.e.}, we define for $f\in\Hom\Q$,
\begin{equation}
(\hat d(f)\psi)(A):=\left\{ \begin{array}{ll}
            \psi(\Ran f)&\mbox{if $\Dom f=A$;}\label{Def:df2}\\
                 0 & \mbox{otherwise,}
                 \end{array}
        \right.
\end{equation}
and with $(\hat d(\star)\psi)(A):=0$ for all $A\in\Ob\Q$.
This scheme can be extended to include quantising
`configuration' variables---{\em i.e.}, real-valued
functions $\beta$ on $\Ob\Q$---by defining
\begin{equation}
        (\hat\beta\psi)(A):=\beta(A)\psi(A)\label{Def:beta1}
\end{equation}
for any $\beta:\Ob\Q\rightarrow\mathR$.

By these means we obtain a simple quantum model. But,
several significant problems can be seen already. For
example, no inner product has been specified on the state
functions; we shall discuss this question shortly.
However, the main problem is that this representation of
$\Sem\Q$ fails to separate arrows with the same domain
and range. This is because if $f$ and $g$ are two such
arrows then the action of $\Sem\Q$ on $\Ob\Q_+$ in
\eq{Def:taufA} is such that $\tau_f(A)=\tau_g(A)$ for all
$A\in\Ob\Q_+$.

This problem could be addressed using similar methods to
those adopted later in the context of arrow fields.
However, we will not follow this path here since, anyway,
the definition \eq{Def:df}, or \eq{Def:df2}, of the
operator $\hat d(f)$ has some peculiar features. In
particular, $\hat d(f)$ annihilates any function whose
support lies in the complement of the singleton set
$\{\Dom{f}\}$, which is rather draconian. It seems more
natural to define an operator
\begin{equation}
(\hat a(f)\psi)(A):=\left\{ \begin{array}{ll}
            \psi(\Ran f)&\mbox{\ if $\Dom f=A$;}
                                    \label{Def:afposs}\\
                 \psi(A) & \mbox{\ otherwise,}
                 \end{array}
        \right.
\end{equation}
which leaves the values of $\psi$ unchanged except on the
object $\Dom f$ on which the arrow $f$ naturally acts.
This would correspond to an action of $\Hom\Q$ on $\Ob\Q$
\begin{equation}
    \tau^\prime_f(A):=\left\{ \begin{array}{ll}
                \Ran f &\mbox{\ if $\Dom f=A$;}
                                    \label{Def:tauprime}\\
                 A & \mbox{\ otherwise,}
                 \end{array}
        \right.
\end{equation}
in contrast to \eq{Def:taufA}.

However, the operators defined by \eq{Def:afposs} do not combine
into themselves. For example, let $f:A\rightarrow B$ and
$g:C\rightarrow D$ be arrows with $C\neq A$ and $A\neq D$. Then
\begin{eqnarray}
(\hat a(g)\hat a(f)\psi)(C)&=&(\hat a(f)\psi)(D)=\psi(D)
\label{agafC}\\
(\hat a(g)\hat a(f)\psi)(A)&=&(\hat a(f)\psi)(A)=\psi(B)
\label{agafA}\\
(\hat a(g)\hat a(f)\psi)(E)&=&\psi(E)\mbox{\ \ for\ all\
$E$\ not\ equal\ to\ $C$\ or\ $A$.}\label{agafE}
\end{eqnarray}
But this is not of the form $\hat a(h)\psi$ for any arrow $h$.

The problem lies in the definition \eq{DefAmp2} of the
combination law on $\Sem\Q$ whereby the partial semigroup
$\Hom\Q$ is transformed into a full semigroup. The
introduction of the additional element $\star$ is a
rather crude device, and distinctly ad hoc. As we shall
now see, there is a far more elegant way of associating a
full semigroup with the set of arrows $\Hom\Q$, and it is
within this framework that the quantisation scheme will
be developed further.

\section{The Monoid of Arrow
Fields}\label{Sec:SemigroupArrowFields}
\subsection{The Idea of an Arrow Field}
The constructions used above are very `local' in object
space. For example, when $\tau_f$ acts on $\Ob\Q_+$, the
only object that is affected is $\Dom f$: the rest are
mapped to the `dustbin' element $\star$ that is appended
to $\Ob\Q$. On the other hand, in the motivating case of
a group $G$ that acts on a manifold $Q$, each element
$g\in G$ acts on every element $q$ of $Q$ (of course,
this includes the case where $g$ leaves $q$ fixed),
without the need to append anything to $Q$. This suggests
that it would be profitable to drop the use of $\star$,
and to seek an alternative structure that better
resembles the typical action of a group on a manifold.

We shall do this by choosing for {\em each\/} object $A$ an arrow
whose domain is $A$ (this could be the identity arrow
$\id_A:A\rightarrow A$), and then act on $A$ with it. Thus we
consider maps $X:\Ob\Q\rightarrow\Hom\Q$ such that, for each
$A\in\Ob\Q$, the domain\footnote{The map $X$ can be viewed as a
cross-section of a bundle on $\Ob\Q$ whose fibre over $A\in\Ob\Q$
is the set of all arrows whose domain is $A$.} of $X(A)$ is $A$;
thus $X(A):A\rightarrow B$ for some $B\in\Ob\Q$. Such a map will
be called an {\em out-arrow field\/}, or just an {\em arrow
field\/}\footnote{Similarly, an {\em in-arrow field\/} is a map
$Y:\Ob\Q\rightarrow\Hom\Q$ such that, for each $A\in\Ob\Q$, the
{\em range\/} of the arrow $Y(A)$ is $A$. Only out-arrow fields
will be used in what follows.}, on $\Q$.

For our purposes, the key property of arrow fields is that they
form a full monoid without needing to append any additional
elements. More precisely, if $X_1$ and $X_2$ are arrow fields, we
construct an arrow field $X_2\&X_1$ by defining, for all
$A\in\Ob\Q$, the arrow $(X_2\&X_1)(A)$ to be the composition of
the arrow $X_1(A)$ with the arrow obtained by evaluating $X_2$ on
the range of $X_1(A)$:
\begin{equation}
    (X_2\&X_1)(A):=X_2(\Ran{X_1(A)})\circ X_1(A).
\end{equation}
Put more simply, if $X_1(A):A\rightarrow B$, then
\begin{equation}
    (X_2\&X_1)(A):=X_2(B)\circ X_1(A)
\end{equation}
as summarised in the diagram $A\ \mapright{X_1(A)}\ B\
\mapright{X_2(B)}\ C$.

To prove associativity, it is helpful to use the diagram
\begin{equation}
    A\ \mapright{X_1(A)}\ B\ \mapright{X_2(B)}\ C
        \ \mapright{X_3(C)}\ D.
\end{equation}
Then $[X_3\&(X_2\&X_1)](A)$ is the arrow from $A$ to $D$
given by
\begin{eqnarray}
    [X_3\&(X_2\&X_1)](A)&=&X_3[\Ran{(X_2\&X_1)(A)}]\circ
                        (X_2\&X_1)(A)       \nonumber\\
    &=&X_3(C)\circ (X_2\&X_1)(A)            \nonumber\\
    &=& X_3(C)\circ[X_2(B)\circ X_1(A)]
\end{eqnarray}
for all $A\in\Ob\Q$. On the other hand
\begin{eqnarray}
    [(X_3\&X_2)\&X_1](A)&=&(X_3\&X_2)(B)\circ X_1(A)\nonumber\\
            &=&[X_3(C)\circ X_2(B)]\circ X_1(A)
\end{eqnarray}
for all $A\in\Ob\Q$. Thus the arrow-field associativity
property, $X_3\&(X_2\&X_1)= (X_3\& X_2)\&X_1$, follows
from the associativity of arrow composition in the
category $\Q$.

There is also a unit element for the $\&$-algebraic
structure. This is the arrow field $\iota$ defined by
\begin{equation}
        \iota(A):=\id_A
\end{equation}
for all $A\in\Ob\Q$. Thus the set of all arrow fields on
$\Q$ is a full monoid. We will denote it
$\AF\Q$.\footnote{If desired, an `incidence algebra'
$\mathC[\AF\Q]$ can be associated with the monoid $\AF\Q$
in the same way that $\mathC[\Sem\Q]$ is generated by
$\Sem\Q$.}

\subsection{The Action of $\AF\Q$ on $\Ob\Q$}
The definition of an arrow field is such that, for each
object $A$, $X(A)$ is an arrow whose domain is $A$. Thus
we can define an action $\ell$ of the monoid $\AF\Q$ on
the set $\Ob\Q$ by letting $X\in\AF\Q$ transform
$A\in\Ob\Q$ into the range of the arrow $X(A)$:
\begin{equation}
        \ell_X(A):=\Ran X(A).
\end{equation}
In other words, if $X(A):A\rightarrow B$ then $\ell_X(A):=B$.

This defines a genuine monoid action of $\AF\Q$ on $\Ob\Q$, since,
for all $A\in\Ob\Q$,
\begin{eqnarray}
    \ell_{X_2\&X_1}(A)&:=&\Ran[(X_2\&X_1)(A)]=
                            \Ran[X_2(\Ran X_1(A))\circ
                            X_1(A)]     \nonumber\\
    &=&\Ran[{X_2(\Ran{X_1(A)})}]
\end{eqnarray}
whereas
\begin{equation}
    \ell_{X_2}(\ell_{X_1}(A))=\ell_{X_2}(\Ran X_1(A))
                =\Ran[X_2(\Ran{X_1(A)})]
\end{equation}
and hence, for all $X_1,X_2\in\AF\Q$,
\begin{equation}
    \ell_{X_2}\circ\ell_{X_1}=\ell_{X_2\&X_1}
                    \label{ellmonoidlaw}
\end{equation}
as required.

Note that a significant difference between this action
and that of a Lie group $G$ on a manifold $Q$ is that the
{\em same\/} group element acts at each point $q\in Q$,
whereas, in the arrow-field action, it is not $X$ as a
whole, but rather the arrow $X(A)$ which acts at
$A\in\Ob\Q$, and this arrow is arbitrary for each $A$.
Thus the arrow-field transformations are more like the
action on $Q$ of the full diffeomorphism group ${\rm
Diff}(Q)$ than that of the finite-dimensional subgroup
$G$. If one wanted to emulate the familiar group case
more closely it would be necessary to relate the arrows
at different objects in some way. For most categories
$\Q$ there is no obvious way of doing this since the
different objects in the category are frequently very
different from each other.\footnote{One exception is when
a manifold $Q\simeq G/H$ is regarded as a category in the
way discussed later in Section
\ref{SubSec:IntroducingMultiplier}. In this case, we can
define the special arrow fields $X^g$, $g\in G$, as
$X^g(q):q\rightarrow gq$ for all $q\in Q$.} There are
some specific examples of `constant' arrow fields in
\cite{IshQCT2_03}.

\subsection{The Special Arrow-Fields $X_f$}
An arrow-field $X$ assigns an arbitrary arrow $X(A)$ to
each object $A$ subject only to the requirement that
$\Dom X(A)=A$. However, a particularly simple choice of
$X$ is when all but one of the arrows $X(A)$,
$A\in\Ob\Q$, is the identity $\id_A$. More precisely, for
each arrow $f\in\Hom\Q$, an arrow field $X_f$ can be
defined by
\begin{equation}
    X_f(A):=\left\{ \begin{array}{ll}
            f &\mbox{\ if $\Dom f=A$;}
                                    \label{Def:Xf}\\
                 \id_A & \mbox{\ otherwise,}
                 \end{array}
        \right.
\end{equation}
for all $A\in\Ob\Q$.

The action of $X_f$ on the set $\Ob\Q$ is
\begin{equation}
    \ell_{X_f}(A)=\left\{ \begin{array}{ll}
                \Ran f &\mbox{\ if $\Dom f=A$;}
                                    \label{Def:ellf}\\
                 A & \mbox{\ otherwise,}
                 \end{array}
        \right.
\end{equation}
which should be contrasted with the definition of $\tau_f$ in
\eq{Def:taufA}. The transformation \eq{Def:ellf}  was anticipated
in \eq{Def:tauprime}.

A natural extension is to pick any finite set of elements
$f_1,f_2,\ldots,f_n\in\Hom\Q$, each of which has a
different domain from the others. We can then define the
arrow-field
\begin{equation}
    X_{f_1,f_2,\ldots,f_n}(A):=\left\{ \begin{array}{ll}
            f_i &\mbox{\ if $\Dom f_i=A$, $i=1,2,\ldots,n$;}
                                    \label{Def:Xfn}\\
                 \id_A & \mbox{\ otherwise.}
                 \end{array}
        \right.
\end{equation}
Arrow fields of this type have finite support, where the
{\em support\/} of an arrow field is defined to be the
set of all objects $A\in\Ob\Q$ such that $X(A)\neq\id_A$.

The collection of all arrow fields of finite support is a
submonoid of $\AF\Q$, and is likely to play an important
role in a deeper analysis of the quantum theory. From a
mathematical perspective, its role could perhaps be
compared with that of the group of gauge transformations
of {\em compact\/} support in a normal gauge theory
(although it must be emphasised that the physical
significance of $\AF\Q$ is not the same as that of a
standard gauge group).

One might anticipate that the set of arrow fields of finite
support can be constructed by taking the $\&$ product of arrow
fields of the type $X_f$. This is true, but the order of the
elements in the product is important. For example, consider a pair
of arrows $f:A\rightarrow B$ and $g:C\rightarrow D$ with $A\neq
C$. Then, if $B\neq C$,
\begin{equation}
            X_f\&X_g=X_g\&X_f=X_{f,g}.
\end{equation}
On the other hand, if $B=C$, so that we have the chain
$A\mapright{f}B\mapright{g}D$, and if $A\neq B$, then
$X_f\&X_g=X_{f,g}$ but $X_g\&X_f=X_{{g\circ f},g}\neq X_{f,g}$.

Finally, note that if $\Dom f\neq \Ran f$ then
\begin{equation}
        X_f\&X_f =X_f,
\end{equation}
so these are {\em idempotent\/} elements of the monoid $\AF\Q$. On
the other hand, if $\Dom f=\Ran f$ then $X_f\&X_f=X_{f\circ f}$.

\section{Arrow-Field Quantum Theory}\label{Sec:AFQT}
\subsection{Quantisation on a Manifold $Q$}
\label{SubSec:QuantManQ} To motivate what follows, consider first
a classical system whose configuration space is a manifold $Q$ on
which there is a transitive left action by a Lie group $G$ with
$Q\simeq G/H$, where $H$ is a closed subgroup of $G$. Thus, to
each $g\in G$ there is a diffeomorphism $\tau_g:Q\rightarrow Q$
with
\begin{equation}
        \tau_{g_2}\circ\tau_{g_1}= \tau_{g_2g_1}
                        \label{Gtauproduct}
\end{equation}
for all $g_1,g_2\in G$.

The classical state space is the cotangent bundle $T^*Q$, and the
quantisation scheme advocated in \cite{Isham84} involves finding
the smallest finite-dimensional group of symplectic
transformations that acts transitively on $T^*Q$. This is a
semi-direct product $G\times_\tau W$ (the `$\tau$' denotes the
action of $G$ on $Q$) where $W$ (a finite-dimensional, linear
subspace of $C^\infty(Q)$) is the dual of the smallest vector
space that carries a linear representation of $G$ with a $G$-orbit
that is equivariantly diffeomorphic to $G/H$. Induced
representation theory \cite{Mac78} shows that the main class of
unitary irreducible representations of $G\times_\tau W$ is given
by vector bundles over this orbit, in which the vector-space fibre
carries an irreducible representation of $H$.\footnote{There may
also be `atypical' representations in which the vector bundle is
over an orbit that is not diffeomorphic to $Q$.}

Note that $G\times_\tau W$ is a {\em finite\/}-dimensional
subgroup of the (infinite-dimensional) group of symplectic
transformations, ${\rm Diff}(Q)\times_d C^\infty(Q)$, of $T^*Q$
(where $d$ denotes the action on $Q$ of the diffeomorphism group,
${\rm Diff}(Q)$, of $Q$). Many of the representations of
$G\times_\tau W$ extend to the group $G\times_\tau C^\infty(Q)$,
and some of these extend to ${\rm Diff}(Q)\times_d C^\infty(Q)$.
However, the general representation theory of the
infinite-dimensional group ${\rm Diff}(Q)\times_d C^\infty(Q)$ is
far more complicated, and incomplete, than that of its
finite-dimensional subgroup $G\times_\tau W$. Unfortunately, if
the manifold $Q$ is {\em not\/} a homogeneous space $G/H$, then
usually one has to fall back on using ${\rm Diff}(Q)\times_d
C^\infty(Q)$.

It would be good to develop a complete analogy of this
scheme for a general small category $\Q$. However, this
involves finding an appropriate analogue of symplectic
geometry, which is not obvious. Here, we will adopt a
more heuristic approach in which we start by thinking of
state vectors as being merely complex-valued functions on
$\Ob\Q$, and then see where this leads in the
construction of the analogue of the quantisation group
$G\times_\tau W$; or, perhaps more precisely, of the
group ${\rm Diff}(Q)\times_d C^\infty(Q)$.

In the manifold case, if there is some $G$-invariant measure $\mu$
on $Q$, then a representation of $G$ on the Hilbert space
$L^2(Q,d\mu)$ of complex-valued functions on $Q$ can be defined by
\begin{equation}
      (\hat U(g)\psi)(q):=\psi(\tau_{g^{-1}}(q)),
            \label{Def:UgRep}
\end{equation}
so that $\hat U(g_2)\hat U(g_1)=\hat U(g_2g_1)$ for all
$g_1,g_2\in G$. If $\mu$ is invariant under the action of
$G$ on $Q$ then this representation is unitary. Note that
if $\hat U(g)$ is defined instead as
\begin{equation}
    (\hat U(g)\psi)(q):=\psi(\tau_g(q)) \label{Def:UgAntiRep}
\end{equation}
then $\hat U(g_2)\hat U(g_1)=\hat U(g_1g_2)$ for all
$g_1,g_2\in G$, {\em i.e.,} we get an {\em
anti}-representation of $G$.

When $Q\simeq G/H$, the representation \eq{Def:UgRep} can be used
as the basis for a simple quantisation of the system. This
involves defining operator representations of configuration
variable functions $\beta\in C^\infty(Q)$, by
\begin{equation}
            (\hat\beta\psi)(q):=\beta(q)\psi(q)
\end{equation}
which can be exponentiated to give the unitary operators
\begin{equation}
          (\hat V(\beta)\psi)(q):=e^{-i\beta(q)}\psi(q).
\end{equation}
Together, $\hat U(g)$ and $\hat V(\beta)$ satisfy the
relations
\begin{eqnarray}
    \hat U(g_1)\hat U(g_2)&=&\hat U(g_1g_2)\label{CCRU}\\
    \hat V(\beta_1)\hat V(\beta_2)&=&\hat
    V(\beta_1+\beta_2)\label{CCRV}\\
\hat U(g)\hat V(\beta)&=&\hat
V(\beta\circ\tau_{g^{-1}})\hat U(g)
    \label{CCRUV}
\end{eqnarray}
where
$\beta\circ\tau_{g^{-1}}(q):=\beta(\tau_{g^{-1}}(q))$. If
the definition \eq{Def:UgAntiRep} is used instead of
\eq{Def:UgRep}, we get the relations
\begin{eqnarray}
    \hat U(g_2)\hat U(g_1)&=&\hat U(g_1g_2)\label{CCRAU}\\
    \hat V(\beta_1)\hat V(\beta_2)&=&\hat
    V(\beta_1+\beta_2)\label{CCRAV}\\
    \hat U(g)\hat V(\beta)&=&
        \hat V(\beta\circ\tau_{g})\hat U(g).
    \label{CCRAUV}
\end{eqnarray}
For this system, \eqs{CCRU}{CCRUV} (or \eqs{CCRAU}{CCRAUV}) are
the analogue of the (exponentiated) canonical commutation
relations of elementary wave mechanics.\footnote{To be more
precise, this is so when the functions $\beta:Q\rightarrow\mathR$
are restricted to belong to the finite-dimensional subspace
$W\subset C^\infty(Q)$ mentioned earlier. However, a
representation of $G\times_\tau W$ on sections of vector bundles
over $Q$ can be extended to include all $C^\infty$ functions on
$Q$ (modulo the usual subtleties with operator domains).} They
constitute a representation of the subgroup $G\times_\tau
C^\infty(Q)$ of the much larger group ${\rm Diff}(Q)\times_d
C^\infty(Q)$.

\subsection{The Basic Algebra for the Quantum Theory on $\Q$}
\label{SubSec:BasicAlgebraQT} Our first task is to find the
analogue of \eqs{CCRAU}{CCRAUV} for a system whose configuration
space (or history-theory equivalent) is the set of objects $\Ob\Q$
in a small category $\Q$. The key idea is to use the monoid
$\AF\Q$ as an analogue of the group of diffeomorphisms of $Q$.

We start with the simplest approach to constructing a quantum
theory on $\Q$, which is to take the state vectors to be
complex-valued functions on $\Ob\Q$. The action of the monoid
$\AF\Q$ on such functions is (writing $\ell_X(A)$ as $\ell_XA$ for
typographical clarity)
\begin{equation}
        (\hat a(X)\psi)(A):=\psi(\ell_XA)=\psi[\Ran X(A)]
        \label{Def:aX}
\end{equation}
which is like the earlier definition \eq{Def:df} except that there
is no need to augment the set $\Ob\Q$ with the additional element
$\#$.

We have
\begin{eqnarray}
    (\hat a(X_2)\hat a(X_1)\psi)(A)&=&(\hat
    a(X_1)\psi)(\ell_{X_2}A)=\psi(\ell_{X_1}(\ell_{X_2}A))
    =\psi(\ell_{X_1\&X_2}A)\nonumber\\
    &=& (\hat a(X_1\& X_2)\psi)(A)
\end{eqnarray}
where \eq{ellmonoidlaw} has been used. Thus we have an
anti-representation of the monoid $\AF\Q$:
\begin{equation}
        \hat a(X_2)\hat a(X_1)=\hat a(X_1\&X_2)
\end{equation}
for all $X_1,X_2\in\AF\Q$.

If \eq{Def:aX} is applied to the special arrow fields $X_f$
in \eq{Def:Xf} then, defining $\hat a(f):=\hat a(X_f)$, we
get
\begin{equation}
    (\hat a(f)\psi)(A)=\left\{ \begin{array}{ll}
                \psi(\Ran f)&\mbox{\ if $\Dom f=A$;}
                \label{aXf}\\
                 \psi(A) & \mbox{\ otherwise,}
                 \end{array}
        \right.
\end{equation}
as  anticipated in \eq{Def:afposs}. The `closure' problem
that arose earlier in the context of \eqs{agafC}{agafE}
no longer applies since the monoid product of two arrow
fields $X_f$, $X_g$ is itself an arrow field (albeit,
possibly not of the form $X_h$ for any $h\in\Hom\Q$).

One might wonder what the adjoint of an operator $\hat a(X)$ looks
like, but this cannot be answered before putting an inner product
on the state functions, which of course is essential anyway to the
physical interpretation of the theory. However, there is no
obvious way of doing this in general. If $\Ob\Q$ is a finite, or
countably infinite, set  we can define
\begin{equation}
        \braket\phi\psi:=\sum_{A\in\Ob\Q}\phi(A)^*\psi(A),
            \label{ScalProdCount}
\end{equation}
although it would be nice to have some specific physical,
or mathematical, reason for choosing this particular inner
product.

More generally, we need to explore the construction of
`appropriate' measures $\mu$ on $\Ob\Q$ so that we can define
\begin{equation}
    \braket\phi\psi:=\int_{\Ob\Q}d\mu(A)\,\phi^*(A)\psi(A).
                \label{ScalProdMeas}
\end{equation}
The first step is to find fields of measurable sets on $\Ob\Q$,
and the easiest way to do this is if there is a topology on
$\Ob\Q$. For example, in the special case where $\Q$ is a poset
$P$ there are the order topologies on $P$ ({\em i.e.} generated by
the upper or lower sets of $P$) and there are probably analogues
of these on a general small category. However, this is a
complicated issue, and is deferred to a later paper. For the
purposes of the present paper it will be assumed that $\Ob\Q$ is
finite or countable, so that the simple inner product
\eq{ScalProdCount} can be used. This is reasonable since many of
the physically interesting examples do have a countable collection
of objects.

The next step in the construction of our `category quantisation
monoid' is to represent the space of real-valued functions on
$\Ob\Q$ (the `configuration variables') by
\begin{equation}
    (\hat\beta\psi)(A):=\beta(A)\psi(A)\label{Def:beta2}
\end{equation}
as in \eq{Def:beta1}.

The crucial task now is to extract an algebra from the operators
$\hat a(X)$ and $\hat\beta$. To this end, we first compute
\begin{equation}
    (\hat a(X)\hat\beta\,\psi)(A)=(\hat\beta\psi)(\ell_XA)=
    \beta(\ell_XA)\psi(\ell_XA)\label{aXbeta}
\end{equation}
while
\begin{equation}
    (\hat\beta\,\hat a(X)\psi)(A)=\beta(A)(\hat
    a(X)\psi)(A)=\beta(A)\psi(\ell_XA)
\end{equation}
which implies that
\begin{equation}
    (\widehat{\beta\circ\ell_X}\hat a(X)\psi)(A)=
    \beta(\ell_XA)\psi(\ell_XA)     \label{betaellXa}
\end{equation}
where $\widehat{\beta\circ\ell_X}$ is defined by
$(\widehat{\beta\circ\ell_X}\psi)(A):=\beta(\ell_XA)\psi(A)$ for
all $A\in\Ob\Q$. From \eq{aXbeta} and \eq{betaellXa} we obtain the
relation
\begin{equation}
        \hat a(X)\,\hat\beta=\widehat{\beta\circ\ell_X}\,\hat
        a(X)            \label{aXb=blXaX}
\end{equation}
for all $X\in\AF\Q$ and functions
$\beta:\Ob\Q\rightarrow\mathR$.

Next we introduce the unitary operator $\hat
V(\beta):=\exp{-i\hat\beta}$ which satisfies $\hat V(\beta_1)\hat
V(\beta_2)=\hat V(\beta_1+\beta_2)$ for all functions $\beta_1$
and $\beta_2$. Finally, putting these relations together, we get
\begin{eqnarray}
    \hat a(X_2)\hat a(X_1)&=&\hat a(X_1\&X_2)\label{CQAaa}\\
    \hat V(\beta_1)\hat V(\beta_2)&=&\hat
    V(\beta_1+\beta_2)                  \label{CQAVV}\\
    \hat a(X)\hat V(\beta)&=&\hat V(\beta\circ\ell_X)\hat
    a(X)                                \label{CQAaV}
\end{eqnarray}
which should be viewed as the category analogue of
\eqs{CCRAU}{CCRAUV}. In the manifold case of
\eqs{CCRAU}{CCRAUV} we have a representation of the group
$G\times_\tau  C^\infty(Q)$. In the category case of
\eqs{CQAaa}{CQAaV} we have a representation of the
semi-direct product $\AF\Q\times_\ell F(\Ob\Q,\mathR)$ of
the monoid $\AF\Q$ with the vector space
$F(\Ob\Q,\mathR)$ of all real-valued functions on
$\Ob\Q$. In what follows, $\AF\Q\times_\ell
F(\Ob\Q,\mathR)$ will be called the `category
quantisation monoid'.

Note, however, that in the manifold case, the functions
$\beta:Q\rightarrow \mathR$ are not totally arbitrary. At
the very least, they are required to be measurable with
respect to the natural $\sigma$-algebra of sets
associated with the topology on $Q$; and one may well
wish to restrict them to be $C^\infty$. However, no
analogous structure has yet been placed on $\Ob\Q$, and
therefore, as things stand, the only option is to include
all real-valued functions on $\Ob\Q$.

Modulo this caveat, the central idea of the proposed
quantum scheme is that the possible quantum theories on
$\Q$ are given by the different faithful, irreducible
representations of the category quantisation monoid
$\AF\Q\times_\ell F(\Ob\Q,\mathR)$. Each such
representation will satisfy the relations in
\eqs{CQAaa}{CQAaV}. However, some important questions
arise when comparing these with the analogous relations
\eqs{CCRAU}{CCRAUV} for the case where the configuration
space is a manifold $Q\simeq G/H$. For example, the group
$G$ acts {\em transitively\/} on $Q$, and this is an
important requirement in proving the irreducibility of
the representation of the quantisation group
$G\times_\tau C^\infty(Q)$. It is necessary to explore
the analogue for the action of the monoid $\AF\Q$ on
$\Ob\Q$. This issue will be discussed briefly in Section
\ref{SubSec:QuestionIrreducibility}, and is examined in
more detail in a later paper.

Note that the operators $\hat U(g)$, $g\in G$, in
\eqs{CCRAU}{CCRAUV} are unitary, but this will not be the
case for the operators $\hat a(X)$, irrespective of the
choice of the measure $\mu$ on $\Ob\Q$. Indeed, although
it is natural to view an arrow as the analogue of
momentum---in the sense that it transforms one object to
another---objects in a category of structured sets are
typically very different from each other and, in this
sense, $\hat a(X)$ is a type of creation or annihilation
operator. In Section \ref{SubSec:Adjointa(X)} we shall
see how this works in specific examples.

\subsection{Introducing a Multiplier}\label{SubSec:UsingMult}
\subsubsection{The Basic Ideas}\label{SubSec:IntroducingMultiplier}
In the context of arrow fields, we shall say that two arrows $f,g$
are {\em separated\/} in the quantum theory if $\hat
a(X_f)\neq\hat a(X_g)$. In this respect, the representation of the
category quantisation monoid constructed above is inadequate since
it fails to separate arrows that have the same domain and range:
indeed, if $f,g$, are any two such arrows then $\hat
a(X_f)\psi=\hat a(X_g)\psi$ for all states $\psi$. In particular,
it cannot represent any of the internal structure of the objects
in the category as reflected in the sets $\Hom{A,A}$, $A\in\Ob\Q$.
To get such a separation, the crucial step is to refine the
quantum scheme by letting the state functions $\psi$ take their
values in a Hilbert space $\K$ that is larger than $\mathC$.

To motivate what follows, we note first that a system with a
configuration manifold $Q\simeq G/H$, where $G$ is a Lie group,
can be viewed as a special example of this categorial structure.
Specifically: let $\Q$ be the category whose objects are the
points in $Q$, and whose arrows from $q_1\in Q$ to $q_2\in Q$ are
defined to be the group elements $g\in G$ such that $q_2=gq_1$,
where $gq$ denotes the point in $Q$ obtained by acting on $q\in Q$
with $g\in G$ ({\em i.e.}, $gq:=\tau_g(q)$). Thus
\begin{equation}
    \Hom{q_1,q_2}:=\{g\in G\mid q_2=gq_1\}.\label{Homq1q2}
\end{equation}
Composition of group elements regarded as arrows\footnote{In this
example, an arrow field is defined by a function $X:Q\rightarrow
G$. Since we are dealing with manifolds, it would be natural to
require this function to be smooth.} is just the group product.
Thus if $g_1:q_1\rightarrow q_2$ ({\em i.e.}, $q_2=g_1q_1$) and
$g_2:q_2\rightarrow q_3$ ({\em i.e.}, $q_3=g_2q_2$), then we
define $g_2\circ g_1:q_1\rightarrow q_3$ as $g_2g_1:q_1\rightarrow
q_3$. The associativity of the composition of arrows follows from
the associativity of the group product\footnote{This is a
generalisation of the well-known fact that a group can be regarded
as a category with a single object, and whose arrows are the group
elements. In fact, since each arrow is invertible, the category
associated with $Q$ is a {\em groupoid\/}.}. In particular,
\eq{Homq1q2} gives
\begin{equation}
   \Hom{q,q}=\{g\in G\mid q=gq\}=G_q\label{Homqq}
\end{equation}
where $G_q$ denotes the `little group' (or stability group) of the
$G$-action at the point $q\in Q$.

Now suppose that $g_1,g_2$ be arrows with the same domain
and range, so that $g_1:q_1\rightarrow q_2$ and
$g_2:q_1\rightarrow q_2$ for some $q_1,q_2\in Q$. Thus
$q_2=g_1q_1$ and $q_2=g_2q_1$, so that
$q_1=g_1^{-1}g_2q_1$, and hence $g_1^{-1}g_2$ belongs to
the stability group $G_{q_1}$, which is isomorphic to
$H$. If we denote $h:=g_1^{-1}g_2 \in G_{q_1}$, then
$g_2=g_1h$; or, in arrow language, $g_2=g_1\circ h$ where
$h\in\Hom{q_1,q_1}$. Thus, to separate the arrows $g_1$
and $g_2$ (with domain $q_1$) it suffices that
$G_{q_1}\simeq H$ be represented faithfully on $\K$. This
is because, if $R(g)$ denotes the representation of $g\in
G$, then $R(g_2)=R(g_1)R(h)$, and hence
$R(g_2)R(g_1)^{-1}=R(h)$, which, for $h\neq e$ (the
identity element in $G_{q_1}$), will not equal the unit
operator if the representation of $H$ is faithful.

However, in a general small category $\Q$, if
$f_1,f_2\in\Hom{A,B}$ this does not imply the existence
of an arrow $\alpha:A\rightarrow A$ such that
$f_2=f_1\circ \alpha$, or an arrow $\beta:B\rightarrow B$
such that $f_2=\beta\circ f_1$, or even a pair of arrows
$\alpha:A\rightarrow A$, $\beta:B\rightarrow B$ such that
$f_2=\beta\circ f_1\circ\alpha$. Nevertheless, the arrows
that need to be distinguished certainly {\em include\/}
those in the sets $\Hom{A,A}$, $A\in\Ob\Q$, and these are
generally object-dependent. This suggest strongly that,
in general, $\K$ must be {\em object dependent\/}.

In the manifold case when $G$ acts on $Q$,  the standard
procedure \cite{Kir76} for finding group representations
using $\K$-valued functions requires the introduction of
a family of linear maps $m(g,q):\K\rightarrow\K$, $g\in
G$, $q\in Q$, (a so-called `multiplier') and then
defining $ (\hat U(g)\psi)(q)=m(g,q)\psi(gq)$. Therefore,
in the case of a general small category, we are led to
consider a family of Hilbert spaces $\K[A]$, $A\in\Ob\Q$,
with a multiplier, $m(X,A)$, that is a linear map from
$\K[\ell_XA]$ to $\K[A]$.

To summarise: we take a bundle of Hilbert spaces
$\bigcup_{A\in\Ob\Q}\K[A]$ over $\Ob\Q$, whose
cross-sections are to be identified as the quantum
states. For a specific measure $\mu$ on $\Ob\Q$, the
inner product is
\begin{equation}
    \braket\phi\psi:=
        \int_{\Ob\Q}d\mu(A)\,\langle\phi(A),\psi(A)\rangle_{\K[A]}
                \label{Def:innerproduct}
\end{equation}
where $\langle\phi(A),\psi(A)\rangle_{\K[A]}$ denotes the
inner product in the Hilbert space $\K[A]$. The
arrow-field operator is defined as
\begin{equation}
    (\hat
    a(X)\psi)(A):=m(X,A)\psi(\ell_XA)\label{Def:aXmXA}
\end{equation}
where $\ell_X(A):=\Ran{X(A)}$, and
\begin{equation}
    m(X,A):\K[\ell_XA]\rightarrow\K[A]\label{Def:mXA}
\end{equation}
is a linear map from $\K[\ell_XA]$ to $\K[A]$. The
configuration variables are real-valued functions $\beta$
on $\Ob\Q$, and are represented (in exponentiated form)
by the unitary operators
\begin{equation}
    (\hat V(\beta)\psi)(A):=e^{-i\beta(A)}\psi(A).
    \label{Def:Vbetamcase}
\end{equation}

Using \eqs{Def:aXmXA}{Def:Vbetamcase}, it is easy to
check that \eq{CQAaV} is satisfied, and of course
\eq{CQAVV} remains unchanged. However,  to satisfy
\eq{CQAaa} certain conditions must be imposed on the
multipliers. Specifically, we have
\begin{equation}
(\hat a(Y\&X)\psi)(A)=m(Y\&X,A)\psi(\ell_{Y\&X}A)
                =m(Y\&X,A)\psi(\ell_Y(\ell_XA))
\end{equation}
while
\begin{eqnarray}
    (\hat a(X)\hat a(Y)\psi)(A)&=&m(X,A)(\hat
              a(Y)\psi)(\ell_XA)\nonumber       \\
    &=&m(X,A)m(Y,\ell_XA)\psi(\ell_Y(\ell_XA)).
\end{eqnarray}
Hence the required condition is
\begin{equation}
    m(Y\&X,A)=m(X,A)m(Y,\ell_XA)   \label{multconds}
\end{equation}
for all arrow fields $X,Y$ and all $A\in\Ob\Q$.

Note that if $X$ and $Y$ are such that $X(A):A\rightarrow
A$ and $Y(A):A\rightarrow A$, then \eq{multconds} gives
\begin{equation}
            m(Y\&X,A)=m(X,A)m(Y,A)  \label{RepHom(AA)}
\end{equation}
which corresponds to an anti-representation of the monoid
$\Hom{A,A}$ on the Hilbert space $\K[A]$.

The goal in choosing the individual Hilbert spaces
$\K[A]$ is to distinguish different arrows between the
same objects. Hence, in particular,  the representation
\eq{RepHom(AA)} of the monoid $\Hom{A,A}$ on $\K[A]$ must
be faithful for all $A\in\Ob\Q$. Of course, this may not
be sufficient to distinguish arrows between {\em
different\/} objects.

It is possible to introduce multipliers even when the
state vectors are only complex-valued functions on
$\Ob\Q$. A multiplier would then be a family of complex
numbers $m(X,A)$, $X\in\AF\Q$, $A\in\Ob\Q$, satisfying
the consistency conditions in \eq{multconds}. However,
such an addition to the simple quantum theory is unlikely
to help with the problem of separating arrows with the
same domain and range. For example, in a category of
sets, the monoid $\Hom{A,A}$ is non-abelian for any set
$A$ with more than one element: as such, it cannot be
represented faithfully with multipliers that are complex
numbers.

\subsubsection{Equivalent and Inequivalent Multipliers}
Let us now discuss briefly the question of when two multipliers
give rise to unitarily equivalent representations of the category
quantisation monoid. Consider a function $A\mapsto L(A)$ which to
each $A\in\Ob\Q$ associates an invertible linear operator in the
Hilbert space $\K[A]$ ({\em i.e.}, $L(A)\in GL(\K[A])$). Let
$B:=\Ran X(A)$ and $C:=\Ran{Y(\Ran X(A))}$ (so that
$X(A):A\rightarrow B$ and $Y(B):B\rightarrow C$) and define a
linear map $m_L(X,A):\K[B]\rightarrow\K[A]$ by
\begin{equation}
        m_L(X,A):=L(A)m(X,A)L(B)^{-1}\label{Def:mL}
\end{equation}
for all $X\in\AF\Q$ and $A\in\Ob\Q$. Then
\begin{equation}
    m_L(Y\&X,A)=L(A)m(Y\&X,A)L^{-1}(C),
\end{equation}
whereas
\begin{eqnarray}
m_L(X,A)m_L(Y,B)&=&L(A)m(X,A)L(B)^{-1}\,L(B)m(Y,B)L(C)^{-1}
                            \nonumber\\
        &=&L(A)m(X,A)m(Y,B)L(C)^{-1}=L(A)m(Y\&X,A)L(C)^{-1}
                            \nonumber\\
        &=& m_L(Y\&X,A).
\end{eqnarray}
Hence \eq {multconds} is satisfied, and so
$m_L(X,A):=L(A)m(X,A)L(B)^{-1}$ is also a multiplier.

If the operators $L(A)$ are unitary for all $A\in\Ob\Q$
({\em i.e.}, $L(A)\in U(\K[A])$), then the representation
of the monoid $\AF\Q$   defined by $(\hat
a(X)\psi)(A):=m_L(X,A)\psi(\ell_XA)$ is clearly unitarily
equivalent to the one obtained using $m(X,A)$. This
suggests that there is a family of inequivalent
multipliers classified by functions $A\mapsto
GL(\K[A])/U(\K[A])$.\footnote{Following the nomenclature
used in group theory, we could say that a quantity $m$
satisfying \eq{multconds} is a {\em one-cocycle} of the
monoid $\AF\Q$ in its action on $\Ob\Q$. Furthermore, two
multipliers/one-cocycles that are related as in
\eq{Def:mL} and with all the $L(A)$, $A\in\Ob\Q$ being
unitary, could be said to differ by a {\em
one-coboundary}.}

\subsubsection{The Presheaf Perspective} As things stand\footnote{This
section can be safely ignored at a first reading: a
knowledge of presheafs is not essential for the theory
being developed.}, the linear map $m(X,A):\K[\Ran
X(A)]\rightarrow\K[A]$ could depend on the values of the
arrow-field $X$ at objects $B$ other than $A$. However,
such a `non-local' property seems unnatural, and from now
on we will suppose that the dependence of $m(X,A)$ on
$X\in \AF\Q$ and $A\in\Ob\Q$ is via the arrow $X(A)$
only. Hence
\begin{equation}
        m(X,A)=\kappa(X(A))
\end{equation}
for some $\kappa(X(A)):\K[\Ran X(A)]\rightarrow \K[A]$.

The consistency conditions \eq{multconds} on the multiplier
become
\begin{equation}
        \kappa[(Y\&X)(A)]=\kappa[X(A)]\kappa[Y(B)]
        \label{kYX}
\end{equation}
where $B,C\in\Ob\Q$ are such that $X(A):A\rightarrow B$
and $Y(B):B\rightarrow C$. But $(Y\&X)(A)=Y(B)\circ
X(A)$, and so \eq{kYX} becomes
\begin{equation}
        \kappa[Y(B)\circ X(A)]=\kappa[X(A)]\kappa[Y(B)].
\label{kYX2}
\end{equation}

However, given any arrow $f\in\Hom\Q$, there is at least
one arrow field $X$ such that $X(\Dom f)=f$ (for example,
$X_f$ defined in \eq{Def:Xf} has this property). Thus a
multiplier $\kappa$ satisfying \eq{kYX} determines linear
maps $\kappa(f):\K[\Ran f]\rightarrow\K[\Dom f]$, for all
$f\in\Hom\Q$. From \eq{kYX2}, these satisfy the
conditions
\begin{equation}
   \kappa(g\circ f)=\kappa(f)\kappa(g)
   \label{kmatch}
\end{equation}
for all $f,g\in\Hom\Q$ such that $\Ran f=\Dom g$ (so that
$g\circ f$ is defined). Conversely, any family of maps
$\kappa(f):\K[\Ran f]\rightarrow\K[\Dom f]$,
$f\in\Hom\Q$, that satisfies \eq{kmatch}, gives rise to a
multiplier defined by $m(X,A):=\kappa(X(A))$.

Such a family of maps $\kappa(f)$, $f\in\Hom\Q$, corresponds
precisely to a {\em presheaf\/}\footnote{Here, a presheaf is
defined as a {\em contravariant\/} functor from $\Q$ to the
category of sets (in our case, Hilbert spaces).} of Hilbert spaces
on $\Q$, and this is the most elegant language in which to
summarise what we have done so far. Namely, we are constructing
representations of the category quantisation monoid, and hence
satisfying, \eqs{CQAaa}{CQAaV} in the following way:
\begin{enumerate}
\item Find a presheaf $\K$ of Hilbert spaces $A\mapsto \K[A]$
over $\Q$. Thus for any arrow $f:A\rightarrow B$ there is a linear
map $\kappa(f):\K[B]\rightarrow\K[A]$, and these linear maps
satisfy the `coherence conditions' that if
$A\mapright{f}B\mapright{g}C$ then
$\K[A]\mapleft{\kappa(f)}\K[B]\mapleft{\kappa(g)}\K[C]$ with
$\kappa(f)\kappa(g)=\kappa(g\circ f):\K[C]\rightarrow\K[A]$.

\item Define the quantum states to be
cross-sections of the corresponding bundle of Hilbert
spaces $\bigcup_{A\in\Ob\Q}\K[A]$; thus $\psi(A)\in\K[A]$
for all $A\in\Ob\Q$. The inner product is
\eq{Def:innerproduct} for some measure $\mu$ on $\Ob\Q$.

\item An arrow field $X\in\AF\Q$ is represented by
the operator
\begin{equation}
        (\hat a(X)\psi)(A):=\kappa(X(A))\psi(\ell_XA).
\end{equation}
Functions $\beta:\Ob\Q\rightarrow\mathR$ are represented
(in exponentiated form) by the unitary operators
\begin{equation}
    (\hat V(\beta)\psi)(A):=e^{-i\beta(A)}\psi(A)
\end{equation}
for all $A\in\Ob\Q$.
\end{enumerate}
The goal is to find a presheaf $\K$ such that the ensuing
representation of the category quantisation monoid is irreducible
and can separate arrows with the same domain and range.

Note that, although a presheaf structure is a fundamental
ingredient in our scheme, it is {\em not\/} the case that
the states $\psi$ are defined as sections (or `global
elements') of this presheaf. Indeed, such a section $\psi$
would satisfy the matching conditions
\begin{equation}
    \psi(A)=\kappa(f)\psi(B)       \label{sectionpresheaf}
\end{equation}
if $f:A\rightarrow B$. This would imply that
\begin{equation}
    (\hat a(X)\psi)(A)=\psi(A)
\end{equation}
for all arrow fields $X$ and objects $A$. This is why the
states are defined to be sections of the {\em bundle\/} of
Hilbert spaces associated with the presheaf, rather than
sections of the presheaf itself: a section of the bundle
does not have to satisfy \eq{sectionpresheaf}.

\subsection{The Adjoint of $\hat a(X)$}
\label{SubSec:Adjointa(X)}
\subsubsection{The Simple Case With no Multipliers}
The next step is to find the adjoints of the operators $\hat
a(X)$, $X\in\AF\Q$. We start with the simple situation in which
there are no multipliers, so that the state vectors are just
complex-valued functions on $\Ob\Q$. We shall also assume
initially that the category $\Q$ contains only a finite number of
objects. Hence, we can use the inner product
\begin{equation}
        \braket{\phi}\psi:=\sum_{A\in\Ob\Q}\phi(A)^*\psi(A)
        \label{InnerProductFiniteQ}
\end{equation}
and then, as usual, for all vectors $\ket\phi,\ket\psi$ we
have
\begin{eqnarray}
        \bra\phi\hat a(X)^\dagger\ket\psi
            &=&\bra\psi a(X)\ket\phi^*\nonumber\\
            &=&\sum_{A\in\Ob\Q}\phi(\ell_XA)^*\psi(A)
\end{eqnarray}
To illustrate what this means let us take a simple example of a
category with five objects $\{A_1,A_2,A_3,B,C\}$, and the
particular arrow field $X$ defined by
\begin{eqnarray}
    X(A_1):A_1&\rightarrow& B\\\label{5elementCat}
    X(A_2):A_2&\rightarrow& B\nonumber\\
    X(A_3):A_3&\rightarrow& B\nonumber\\
    X(B):B&\rightarrow& C    \nonumber\\
    X(C)=\id_C:C&\rightarrow& C.
\end{eqnarray}
Then we have
\begin{equation}
\bra\phi\hat a(X)^\dagger\ket\psi=
        \phi(B)^*(\psi(A_1)+\psi(A_2)+\psi(A_3))+
            \phi(C)^*\psi(B)+\phi(C)^*\psi(C)
\end{equation}
where the last term comes from the fact that $X(C)=\id_C$.

It is clear that, in general, we can write
\begin{equation}
\bra\phi\hat a(X)^\dagger\ket\psi=
    \sum_{B\in\Ob\Q}\sum_{A\in\ell_X^{-1}\{B\}}\phi(B)^*\psi(A)
\end{equation}
where we have been able to sum over all $B\in\Ob\Q$ by allowing
for the fact that $\ell_X^{-1}\{B\}$ may be the empty set for some
objects $B$. Thus we see that
\begin{equation}
    (\hat a(X)^\dagger\psi)(B)=
\sum_{A\in\ell_X^{-1}\{B\}}\psi(A).\label{aXdagPsiB}
\end{equation}
This result can be extended to the case where the set of objects
$\Ob\Q$ is countably infinite, although the usual care will need
to be taken with the domains of the operators $\hat a(X)$ and
their adjoints.

If Dirac notation is used, we write $\psi(A)$ as
$\braket{A}\psi$, in which case the equation $(\hat
a(X)\psi)(A):=\psi(\ell_XA)$ reads $\bra A\hat
a(X)\ket\psi=\braket{\ell_XA}\psi$, and so
\begin{equation}
    \hat a(X)^\dagger\ket{A}=\ket{\ell_X A}.
    \label{aXdagDiracNotation}
\end{equation}
In particular, this shows that, for any object
$A\in\Ob\Q$,  $\hat a(X)^\dagger\ket{A}$ is never
zero\footnote{Of course, this does not exclude the
existence of states $\ket\psi$ for which $\hat
a(X)^\dagger\ket\psi=0$.}. In this restricted sense,
$\hat a(X)^\dagger$ looks like a type of {\em creation\/}
operator.

On the other hand, the equation $(\hat
a(X)^\dagger\psi)(B)= \sum_{A\in\ell_X^{-1}\{B\}}\psi(A)$
becomes
\begin{equation}
    \bra{B}\hat a(X)^\dagger\ket\psi =
        \sum_{A\in\ell_X^{-1}\{B\}}\braket{A}\psi
\end{equation}
and so
\begin{equation}
    \hat a(X)\ket{B}=\sum_{A\in\ell_X^{-1}\{B\}}\ket{A}.
                \label{aXketB}
\end{equation}
In particular, if $B$ is an object that is not the range of any
arrow in the arrow field $X$, then $\ell_X^{-1}\{B\}=\emptyset$,
and hence
\begin{equation}
        \hat a(X)\ket{B}=0.
\end{equation}
Thus, in this restricted sense, $\hat a(X)$ looks like a
type of {\em annihilation operator}.

To illustrate these results concretely, let us return to the
simple category with five objects $\{A_1,A_2,A_3,B,C\}$, and the
arrow field shown in \eq{5elementCat}. Since no arrows in the
arrow field $X$ enter $A_1$, $A_2$, or $A_3$ we have
\begin{equation}
    \hat a(X)\ket{A_1}=\hat a(X)\ket{A_2}=\hat
    a(X)\ket{A_3}=0.
\end{equation}
On the other hand $\ell_X^{-1}\{B\}=\{A_1,A_2,A_3\}$, and so
\begin{equation}
    \hat a(X)\ket B=\ket{A_1}+\ket{A_2}+\ket{A_3}.
\end{equation}
Finally, $\ell_X^{-1}\{C\}=\{B,C\}$ (using $X(C)=\id_C$), which
gives
\begin{equation}
        \hat a(X)\ket{C}=\ket{B}+\ket{C}.
\end{equation}

\subsubsection{The Arrow Field Operators $\hat a(f)$.}
The arrow fields $X_f$, $f\in\Hom\Q$, are particularly
interesting as they generate the arrow fields with finite
support. The operators $\hat a(f):=\hat a(X_f)$ acts as
\begin{equation}
    (\hat a(f)\psi)(A)=\left\{ \begin{array}{ll}
                \psi(\Ran f)&\mbox{\ if $\Dom f=A$;}\\
                 \psi(A) & \mbox{\ otherwise,}
                 \end{array}
        \right.
\end{equation}
and so, in Dirac notation,
\begin{equation}
    \hat a(f)^\dagger\ket{A}=\left\{ \begin{array}{ll}
                \ket{\Ran f}&\mbox{\ if $\Dom f=A$;}
                                        \label{aXfDir}\\
                 \ket{A} & \mbox{\ otherwise}
                 \end{array}
        \right.
\end{equation}
Furthermore, from \eq{aXketB} we see that
\begin{equation}
\hat a(f)\ket{A}=\left\{ \begin{array}{ll}
                \ket{\Dom f}&\mbox{\ if $\Ran f=A$;}
                                        \label{adagXfDir}\\
                 0 & \mbox{\ otherwise.}
                 \end{array}
        \right.
\end{equation}
Note that the operators $\hat a(f)$ and $\hat a(f)^\dagger$
are always bounded, even when the quantum Hilbert space is
infinite dimensional.

\subsubsection{The Situation for a General Measure $\mu$ on
$\Ob\Q$.} For a general measure $\mu$ on $\Ob\Q$, if $\hat
a(X)$ is bounded we have the equation
\begin{equation}
\bra\phi\hat a(X)^\dagger\ket\psi
            =\bra\psi\hat a(X)\ket\phi^*\
            =\int_{\Ob\Q}d\mu(A)\,\phi(\ell_XA)^*\psi(A).
            \label{adaggerXgeneral}
\end{equation}
If $\phi$ is chosen to be the characteristic function
$\chi_S$ of a measurable subset $S$ of $\Ob\Q$,
\eq{adaggerXgeneral} gives\footnote{Of course, to do this
rigourously it is necessary to define precisely what is
meant by integrating over vectors.}
\begin{eqnarray}
    \int_Sd\mu(A)\,(\hat a(X)^\dagger\psi)(A)&=&
    \int_{\Ob\Q}d\mu(A)\,\chi_S(\ell_XA)^*\psi(A)
                                    \nonumber\\
    &=&\int_{\ell_X^{-1} [S]}d\mu(A)\,\psi(A)
                                    \label{adaggmu}
\end{eqnarray}
where $\ell_X^{-1}[S]:=\{A\in\Ob\Q\mid \ell_XA\in S\}$.
Note that if $\Ob\Q$ is a finite set, and if $\mu$ is the
point measure that assigns equal weight $1$ to each object
$A$, and if we chose $S:=\{B\}$, then the result in
\eq{aXdagPsiB} is recovered.

\subsection{Products of $\hat a(X)$ and $\hat a(X)^\dagger$}
\label{SubSec:ProdaXadagX}
\subsubsection{The Operator $\hat a(X)^\dagger\hat a(X)$.}
When applying this quantum theory to specific physical situations,
all important physical quantities must be constructed from the
operators $\hat\beta$, $\hat a(X)$ and $\hat a(X)^\dagger$. For
example, if the theory is being interpreted canonically, then the
Hamiltonian will be particularly important. If the theory is
interpreted as a history theory, then the decoherence function
will be of central importance. Here we shall look briefly at the
quadratic expressions $\hat a(X)^\dagger\hat a(X)$ and $\hat
a(X)\hat a(X)^\dagger$.

First, ignoring possible problems with operator domains,
left multiplying \eq{aXb=blXaX} with $\hat a(X)^\dagger$
gives
\begin{equation}
        \hat a(X)^\dagger\hat a(X)\hat\beta=
\hat a(X)^\dagger\widehat{\beta\circ\ell_X}\hat a(X).
\end{equation}
On the other hand, taking the adjoint of \eq{aXb=blXaX} gives
$\hat\beta\,\hat a(X)^\dagger=\hat
a(X)^\dagger\widehat{\beta\circ\ell_X}$, and right multiplying
this with $\hat a(X)$ gives
\begin{equation}
    \hat\beta\,\hat a(X)^\dagger\hat a(X)= \hat
a(X)^\dagger\widehat{\beta\circ\ell_X}\hat a(X).
\end{equation}
Hence,
\begin{equation}
    [\,\hat a(X)^\dagger\hat a(X),\hat\beta\,]=0
\end{equation}
which, working on the assumption that the algebra generated by the
operators of the form $\hat\beta$ is maximal abelian\footnote{More
precisely, assuming that the space $L^\infty(\Ob\Q,d\mu)$ of $\mu$
essentially-bounded real-valued functions on $\Ob\Q$ is maximal
abelian when considered as an algebra of multiplication operators
on $L^2(\Ob\Q,d\mu)$.}, implies that
\begin{equation}
        (\hat a(X)^\dagger\hat a(X)\psi)(A)=\alpha_X(A)\psi(A)
        \label{aXdagaX}
\end{equation}
for some measurable function
$\alpha_X:\Ob\Q\rightarrow\mathR$.

It is easy to compute $\hat a(X)^\dagger\hat a(X)$ explicitly for
the simple case when $\Q$ has a finite number of objects and the
inner product \eq{InnerProductFiniteQ} is used. We get
\begin{eqnarray}
        (\hat a(X)^\dagger\hat a(X)\psi)(A)&=&
\sum_{C\in\ell_X^{-1}\{A\}}(\hat a(X)\psi)(C)\nonumber\\
            &=&\sum_{C\in\ell_X^{-1}\{A\}}\psi(\ell_XC)
            \label{aXdagaXpsiA}
\end{eqnarray}
or, in Dirac notation, $\hat a(X)^\dagger\hat a(X)\ket{A}=
        \sum_{C\in\ell_X^{-1}\{A\}}\ket{\ell_XC}$. However,
for each $C\in\ell_X^{-1}\{A\}$ we have $\ell_XC=A$, and hence
\begin{equation}
\hat a(X)^\dagger\hat a(X)\ket{A}=
            |\ell_X^{-1}\{A\}|\ket{A}
\end{equation}
where $|\ell_X^{-1}\{A\}|$ denotes the number of elements in the
set $\ell_X^{-1}\{A\}$. As is to be expected, this is consistent
with the general result in \eq{aXdagaX}.\footnote{When there is a
general measure $\mu$ on $\Q$ the calculations are more
complicated. However, it can be shown that if $\hat a(X)$ is a
bounded operator, and if $\mu$ is a finite measure, then the
Radon-Nikodym derivative ${d\ell_X{}_*\mu\over d\mu}$ exists and
is equal to the function $\alpha_X$ in \eq{aXdagaX}. Thus, in
these circumstances, we have
\begin{equation}
    (\hat a(X)^\dagger\hat a(X)\psi)(A)=
    {d\ell_X{}_*\mu\over d\mu}(A)\psi(A)
\end{equation}
everywhere except on a set of $\mu$-measure zero.}

For example, in the model category discussed earlier with the
arrow field $X$ in \eq{5elementCat}, we have
\begin{equation}
\hat a(X)^\dagger\hat a(X)\ket{B}=
            3\ket{B}.
\end{equation}

\subsubsection{The Operator $\hat a(X)\hat a(X)^\dagger$.}
In a similar way, we can compute
\begin{eqnarray}
    (\hat a(X)\hat a(X)^\dagger\psi)(A)&=
    &(\hat a(X)^\dagger\psi)(\ell_XA)\nonumber\\
    &=&\sum_{C\in\ell_X^{-1}\{\ell_X A\}}\psi(C)
    \label{aXaXdagpsiA}
\end{eqnarray}
where $\ell_X^{-1}\{\ell_X A\}$ is the set of all arrows in
the arrow field $X$ whose range is $\ell_XA$. In Dirac
notation this reads
\begin{equation}
    \hat a(X)\hat a(X)^\dagger\ket{A}=
    \sum_{C\in\ell_X^{-1}\{\ell_X A\}}\ket{C}.
\end{equation}

For example, in the model category with five objects
$\{A_1,A_2,A_3,B,C\}$ and the arrow field represented by
\eq{5elementCat}, we have, for $i=1,2,3$,
\begin{equation}
    \ell_X^{-1}\{\ell_X A_i\}=\{A_1,A_2,A_3\}
\end{equation}
and so
\begin{equation}
\hat a(X)\hat a(X)^\dagger\ket{A_i}=
    \ket{A_1}+\ket{A_2}+\ket{A_3}
\end{equation}
for $i=1,2,3$.

\subsection{Including the Multiplier.} The calculations
become more complicated when the state vectors are
Hilbert-space valued with a multiplier $m$. However, the
essence is the same, and here we just quote some of the
results.

The first is that the adjoint $\hat a(X)^\dagger$ is given
by (c.f. \eq{aXdagPsiB})
\begin{equation}
    (\hat a(X)^\dagger\psi)(B)=
        \sum_{A\in\ell_X^{-1}\{B\}}m(X,A)^\dagger\psi(A)
\end{equation}
where $m(X,A)^\dagger:\K[A]\rightarrow\K[\ell_XA]$ is the
adjoint\footnote{These adjoint maps $m(X,A)^\dagger\equiv
\kappa(X(A))^\dagger$ define a {\em covariant\/} functor
from $\Q$ to the category of sets (in our case, Hilbert
spaces).} of the linear map
$m(X,A):\K[\ell_XA]\rightarrow\K[A]$.

When the category $\Q$ has only a finite number of objects, the
operator products $\hat a(X)\hat a(X)^\dagger$ and $\hat
a(X)^\dagger\hat a(X)$ can be readily computed. Thus we have
(c.f.\ \eq{aXaXdagpsiA})
\begin{eqnarray}
    (\hat a(X)\hat a(X)^\dagger\psi)(A)&=
    &m(X,A)[(\hat a(X)^\dagger\psi)(\ell_XA)]\nonumber\\
    &=&\sum_{C\in\ell_X^{-1}\{\ell_X A\}}
        m(X,A)m(X,C)^\dagger\psi(C)
\end{eqnarray}
and (c.f.\ \eq{aXdagaXpsiA})
\begin{eqnarray}
        (\hat a(X)^\dagger\hat a(X)\psi)(A)&=&
    \sum_{C\in\ell_X^{-1}\{A\}}m(X,C)^\dagger
            [(\hat a(X)\psi)(C)]\nonumber\\
    &=&\sum_{C\in\ell_X^{-1}\{A\}}m(X,C)^\dagger
            m(X,C)\psi(\ell_XC)\nonumber\\
    &=&\sum_{C\in\ell_X^{-1}\{A\}}m(X,C)^\dagger
            m(X,C)\psi(A).
\end{eqnarray}

\subsection{The Question of Irreducibility}
\label{SubSec:QuestionIrreducibility} Finally, something
should be said about the irreducibility, or otherwise, of
these representations of the category quantisation
monoid. When quantising a system whose configuration
space is a manifold $Q\simeq G/H$, the corresponding
quantisation group is the semi-direct product
$G\times_\tau W$, and the unitary equivalence classes of
irreducible representations are classified via induced
representation theory in terms of (i) the orbits of $G$
on the dual of $W$, and (ii) the different irreducible
representations of $H$ \cite{Mac78}.

It remains a task for the future to determine a complete
representation theory for the case of a general small category
$\Q$; if, indeed, this is possible. However, in the manifold
analogy, if $Q$ can be decomposed into more than one $G$-orbit,
then there is a corresponding decomposition of the group
representation into a direct sum or direct integral. This, at
least,  should have an analogue in the category case, and so a
natural question is whether $\Ob\Q$ is a single orbit under the
action of $\AF\Q$.

The concept of an `orbit' is more subtle for an action of
a semigroup on a set than it is for a group, and a fuller
discussion of this issue is deferred to a later paper in
this series. However, on looking at the operators $\hat
a(X)$ and $\hat a(X)^\dagger$ as given, for example, in
\eq{aXdagDiracNotation} and \eq{aXketB} it seems natural
to define a subset $O$ of $\Ob\Q$ to be `connected' if
for any pair of objects $A,B\in O$ there exists a finite
collection of objects $\{A_1,A_2,\ldots,A_N\}\subset O$,
with $A_1=A$, $A_N=B$ and such that, for all
$i=1,2,\ldots N-1$, there exists an arrow with domain
$A_i$ and range $A_{i+1}$, or an arrow with range $A_i$
and domain $A_{i+1}$.

Clearly, if $\Ob\Q$ decomposes into a disjoint union of
connected subsets, then the representation of the
category quantisation monoid will decompose in a
corresponding way. Thus a necessary condition for the
representation to be irreducible is that $\Ob\Q$ is
connected in this sense. All the physical examples we
have mentioned so far have this property. However,
connectedness alone is certainly not sufficient to
guarantee irreducibility, and we will return to this
issue later.

\section{Conclusions}\label{Sec:Conclusions}

We have seen how to construct a quantum scheme for a system
whose configuration space (or history equivalent) is the
set of object $\Ob\Q$ in a small category $\Q$. A key
ingredient is the monoid $\AF\Q$ of arrow fields and its
action on $\Ob\Q$. Multiplier representations are needed to
distinguish quantum theoretically between arrows with the
same range and domain. Each such representation can be
expressed in terms of a presheaf of Hilbert spaces over
$\Ob\Q$.

The material in the present paper is only an introduction to what
needs to be done to construct a complete representation theory of
a category quantisation monoid. Many topics remain for further
research, some of the most important of which are the following.

\begin{enumerate}

\item A general question is how much the theory
can be developed in terms of an arbitrary small category
$\Q$, and how much will need to rely on the special
properties of particular categories of physical interest.

\item It would be good to determine some general way of
specifying the Hilbert spaces $\K[A]$, $A\in\Ob\Q$, such
that the ensuing representation of the category
quantisation monoid is both faithful and irreducible. At
the very least, this is likely to require a proper study
of the meaning, and role, of an orbit of the monoid
$\AF\Q$ as it acts on the set $\Ob\Q$. However, it may be
that a full discussion of irreducibility can only be
given in the context of a case-by-case study with
specific categories $\Q$.

\item The classification of inequivalent irreducible
representations of the category quantisation monoid will
involve the choice of a presheaf of Hilbert spaces, and
the choice of the measure $\mu$ used in the inner product
in \eq{Def:innerproduct}. It is necessary therefore to
develop a proper measure theory on the set $\Ob\Q$.
Whether this is feasible for a general small category
$\Q$ is unclear, but even if it is, it seems likely that
the construction of the physically relevant measures will
depend on the details of the category. For example,
Brightwell et al have recently developed a particular
measure theory on a space of causal sets \cite{Bri02a}
\cite{Bri02b}. This was carried out in the context of
constructing a classical stochastic theory of causal
sets, but perhaps these are also the correct measures to
use in the quantum theory as developed in the present
paper?

\item The quantisation of a system whose configuration space is
a manifold $Q\simeq G/H$, uses only the finite-dimensional
subgroup $G$ of the group of all diffeomorphisms, ${\rm Diff}(Q)$,
and the question arises therefore if, in the category case, there
is some submonoid of $\AF\Q$ that still acts `transitively' on
$\Ob\Q$ and which would be a more appropriate entity to use in the
category quantisation monoid. The answer is likely to depend
strongly on the details of the category $\Q$.

\item In the standard quantum theory of a
system whose configuration space is an {\em infinite\/}
dimensional topological vector space $V$ (for example, in a
quantum field theory), the state vectors are typically
functions on the {\em topological dual\/} of $V$ rather
than on $V$ itself. This is closely connected to the theory
of measures on spaces of this type.

    This raises the intriguing question of whether
something like this could happen when quantising on a
category $\Q$. In other words, are there situations in
which an analogue of the topological dual is needed; and
what is the `dual' of the set of objects $\Ob\Q$? The
answer is likely to be closely linked with the problem of
constructing suitable measures on $\Ob\Q$. It is also
related to the issue of whether the quantum scheme should
involve only some linear subspace of the set of all
real-valued functions on $\Ob\Q$. This would be an
analogue of the use of $W\subset C^\infty(Q)$ when
$Q\simeq G/H$.
\end{enumerate}

When applied to categories of space-times (or spaces) the
scheme described above deals with the quantum states of
those structures only. However, in practice, there will
be other degrees of freedom too (for example, matter
fields), and these need to be incorporated at some point.
This could be done by exploiting whatever is known
already about the quantisation of these extra degrees of
freedom, and adjusting the Hilbert spaces $\K[A]$,
$A\in\Ob\Q$, accordingly. The representation of the
category quantisation monoid will then no longer be
irreducible because of the presence of these extra modes.

However, it may be possible to include any extra degrees
of freedom strictly within the category quantisation
scheme by changing the category $\Q$ to get an
appropriately extended category quantisation monoid. For
example, if $\Q$ is a category of topological spaces,
then one might replace the objects (topological spaces)
with the spaces of continuous functions on them, with
appropriate modifications of the arrows. This would give
a type of quantum field theory on a space that is itself
quantised.

Note that, if $\Q$ is a category of manifolds, additional
degrees of freedom could include quantised metric fields.
For example, it would be possible to construct a canonical
theory of quantum gravity (in either the traditional
formalism, or in the newer scheme based on loop variables)
in which the spatial 3-manifold is itself subject to
`quantum fluctuations'. The analogue in a history theory
would be to quantise Lorentzian metrics on a quantised
background space-time manifold. Or the techniques could be
applied to give a version of string theory in which the
manifold in which the strings, or $d$-branes, propagate is
itself the subject of quantum effects.

Finally, note that, when discussing the quantum theory of causal
sets, I have assumed that the  space $\K[c]$ associated with each
causal set $c$ is a standard Hilbert space, in accordance with
normal quantum theory. However, in \cite{Ish03} it is argued that
normal quantum theory is problematic in such a situation because
the use of the continuum real and complex numbers assumes {\em a
priori\/} that the background space and space-time are manifolds,
which is not the case if the space-time is a causal set. This
suggests that each $\K[c]$ should be replaced by something quite
different: in fact, by whatever the analogue is for that specific
causal set $c$ of the Hilbert space of states in normal quantum
theory. It is a task for future research to decide what this may
be, but once the decision is made, the techniques described in the
present paper would be a good starting point to construct a theory
in which the causal sets, and the associated quantum theories, are
themselves subject to `quantum fluctuations'.

These projects are exciting, but it should be emphasised that what
is described in the present paper is only a `tool-kit' for
constructing operator-based models of quantum space-time or space:
it needs a creative leap to use these tools to construct a
physically realistic model of, for example, quantum causal sets.
The key step would be to choose a decoherence functional for the
quantum history theory. This decoherence functional would be
constructed from the operators described in this paper, but new
physical principles are needed to decide its precise form. This is
an important topic for future research.

\section*{Acknowledgements}

\noindent I thank Jeremy Butterfield and Raquel Gracia
for a critical reading of the first draft of this paper.
Support by the EPSRC in form of grant GR/R36572 is
gratefully acknowledged.

\end{document}